\documentclass[preprint,showpacs,preprintnumbers,amsmath,amssymb]{revtex4}
\usepackage{graphicx}
\usepackage{amsfonts}
\usepackage{amssymb}
\usepackage{amsmath}
\usepackage{flafter}
\usepackage{epstopdf}

\providecommand{\Journal}[4] {#1 {\bf#2}, #4 (#3)}
\providecommand{\Journal}[4] {#1 {\bf#2} (#3) #4}
\providecommand{\PRD}{Phys. Rev. D} %
\providecommand{\ApP}{Astropart. Phys.}%
\providecommand{\SKPAWB}{Sitzungsberichte der K$\ddot{\rm o}$niglich Preu$\ss$ischen Akademie der Wissenschaften zu Berlin}%

\begin{document}

\renewcommand{\baselinestretch}{1.25}

\title{Universal entropy bound and discrete space-time}

\newcommand*{\PKU}{School of Physics and State Key Laboratory of Nuclear Physics and
Technology, \\Peking University, Beijing 100871,
China}\affiliation{\PKU}
\newcommand*{\CHEP}{Center for High Energy
Physics, Peking University, Beijing 100871,
China}\affiliation{\CHEP}
\newcommand*{\CHPS}{Center for History and Philosophy of Science, Peking University, Beijing 100871,
China}\affiliation{\CHEP}


\author{Yunqi Xu}\affiliation{\PKU}
\author{Bo-Qiang Ma}\email{mabq@pku.edu.cn}\affiliation{\PKU}\affiliation{\CHEP}\affiliation{\CHPS}



\begin{abstract}
Starting from the universal entropy bounds suggested by Bekenstein
and Susskind and applying them to the black-body radiation
situation, we get a cut-off of space $ \Delta x \geq \chi
l_{\mathrm{P}}$ with $\chi \geq 0.1$. We go further to get a cut-off
of time $ \Delta t \geq \chi l_{\mathrm{P}}/c $, thus, the discrete
space-time structure is obtained. With the discrete space-time, we
can explain the uncertainty principle. Based on the hypothesis of
information theory and the entropy of black holes, we get the
precise value of the parameter $\chi$ and demonstrate the reason why
the entropy bounds hold.
\end{abstract}





\maketitle

\section{Introduction\label{sec:1}}
The generalized second law (GSL) and entropy bound came out with the
development of black hole physics. In 1971, Hawking firstly derived
the area theorem~\cite{Hawking1,Hawking2}, which states that the
area of a black hole event horizon never decreases with time: $
{\mathrm d} A \geq 0 $. Inspired by this idea together with the
second law in thermodynamics, Bekenstein~\cite{BHSL,BHE,GSL} found
that the area of a black hole takes the role of the entropy of the
black hole and suggested that it is proportional to the entropy:
$S_{\mathrm{BH}}=\alpha A $, where the proportional coefficient is
derived by Hawking~\cite{BHEH} to be $\alpha=1/4$. Moreover,
Bekenstein also proposed that the second law of thermodynamics still
holds for the total entropy including the black hole entropy and
matter entropy outside it. This is the generalized second law: $
\Delta S_{\mathrm{total}}\geq 0 $ and $S_{\mathrm{total}} =
S_{\mathrm{matter}} + S_{\mathrm{BH}}$.

In classical general relativity, black holes can only absorb
particles and never emit anything. However, in 1975, by means of
semi-classical calculations of quantum field theory,
Hawking~\cite{PCBH} made a surprising discovery that black holes can
create and emit particles like any other matter. It is just like a
thermodynamic system with temperature $ T = {\kappa} /{2\pi} $, in
which $\kappa$ is the surface gravity of the event horizon. For a
Schwarzschild black hole, the temperature is ${1}/{8\pi M}$, which
is inversely proportional to the mass of the black hole $M$. The
area theorem is violated, however, the generalized second law still
holds in this thermal emission situation.

The universal entropy bound is a direct inference of the generalized
second law. In Section~\ref{sec:2}, we present two different entropy
bounds from Bekenstein and Susskind. They are equivalent to each
other in many situations. Applying these two entropy bounds to the
black-body radiation, we get two constraints for the temperature of
the black-body. To be consistent with these two constraints, we draw
a conclusion that the space has a minimum length scale proportional
to the Planck length. A minimum time interval can also be obtained,
thus, a discrete space-time structure is obtained. In
Section~\ref{sec:3}, we discuss the relation between the discrete
space-time and the Heisenberg uncertainty principle. From
information theory and a minimum volume introduced in
Section~\ref{sec:4}, we derive a precise value of the parameter
$\chi$. An explanation of the entropy bound is given in this
section. Finally we discuss on the space-time discreteness and
summarize the conclusions of our study.

\section{Entropy bound and discrete space-time \label{sec:2}}
By the analysis of the Geroch process, in which a system is dropped
into a black hole from the vicinity of the event horizon, Bekenstein
found that, if the generalized second law still holds, a universal
entropy bound of the system is necessary. We call it the Bekenstein
bound~\cite{UUB}:
\begin{equation}
S_{\mathrm{matter}} \leq 2\pi E R, \label{BekensteinBound}
\end{equation}
in which $S$ is the entropy of the system, $E$ is the energy of it,
and $R$ is the radius of smallest sphere that fits around the matter
system.

Likewise, considering the Susskind process, in which the system
shrinks and is converted to a black hole, Susskind~\cite{TWAH,THPB}
argued that the generalized second law, if applied to the
transformation, yields a spherical entropy bound:
\begin{equation}
S_{\mathrm{matter}} \leq \frac{A}{4},
\end{equation}
where $A$ is a suitably defined area enclosing the matter system. In
the above two inequalities, only black holes can saturate the
entropy bound, that is to say, at the same confined conditions, the
black holes have the biggest entropy among all matter. So the above
two inequalities are appropriate to all matter.

In the situation of a spherical system, both entropy bounds can be
applied. Consider a spherical system filled with black-body
radiation, in which the temperature is $T$ and the radius is $R$.
From the thermodynamics of black-body radiation, we know that the
entropy of the system is:
\begin{equation}
S = \frac{4}{45}{\pi}^2 {T}^3 V = \frac{16}{135}{\pi}^3 R^3 T^3.
\end{equation}
Applying the entropy to the Susskind entropy bound $ S\leq {A}/{4} =
\pi R^2 $, we get a constraint of temperature $T$:
\begin{equation}
T\leq \Big(\frac{135}{16{\pi}^2 R}\Big)^{\frac{1}{3}}.
\end{equation}
The larger the sphere is, the lower temperature the system has.

Similarly, the ratio of entropy-energy of black-body radiation
satisfies:
\begin{equation}
\frac{S}{E} = \frac{4}{3T}.
\end{equation}
Substituting it to the Eq.~(\ref{BekensteinBound}), we get another
constraint:
\begin{equation}
T\geq \frac{2}{3\pi R}.
\end{equation}
The smaller the sphere is, the higher temperature the system has,
just like the previous one. By combining these two inequalities, the
temperature must satisfy:
\begin{equation}
\frac{2}{3\pi R} \leq T \leq \Big(\frac{135}{16{\pi}^2
R}\Big)^{\frac{1}{3}}.
\end{equation}
To reconcile the two sides of the equation, the left side must be
less than the right side, that is:
\begin{equation}
\frac{2}{3\pi R} \leq \Big(\frac{135}{16{\pi}^2
R}\Big)^{\frac{1}{3}}.
\end{equation}

We thus arrive at the following conclusion:
 \begin{equation}
 R\geq \Big(\frac{128}{3645\pi}\Big)^{\frac{1}{2}} l_{\mathrm{P}} \simeq 0.1
 l_{\mathrm{P}},
 \end{equation}
 which means that space can not be infinitesimal. It must have a
 minimum value, which is proportional to the Planck length $l_{\mathrm{P}}=\sqrt{G\hbar/c^3}$. Therefore
 space is discrete rather than continuous. The minimum length scale is the
 cut-off of space, and it may eliminate the ultraviolet limit
 problem and the infinity in the quantum field theory. To be more
 universal, we set the proportional parameter to be $\chi$, that is,
 $\Delta x \geq \chi l_{\mathrm{P}} $ where $\chi \geq 0.1$.

 As for time, we can also get a minimum time interval
 \begin{equation}
\Delta t = \frac{\Delta x}{v} \geq \frac{\Delta x}{c} \geq
\frac{\chi  l_{\mathrm{P}}}{c} = \chi \sqrt{\frac{G\hbar}{c^5}}.
 \end{equation}
 In the above derivation, we adopt the relation $v\leq c$, where $v$
 is the velocity of an ordinary particle and $c$ is the velocity of
 light. If we adopt the natural unit system, {\it i.e.}, we set $G = \hbar = c =
 k = 1$, then $\Delta t \geq \Delta x \geq \chi$. The minimum volume of
 space is $\Delta V \geq \chi^3$.
 Therefore time has a minimum interval, which implies that the motion of a
 particle can not be divided to be infinitesimal but should be quantized. It is much like
 the transition from one quantum state to another one. As time is
 not infinitesimal, the time paradox put forward by Zeno can be solved easily.

\section{The Heisenberg uncertainty relation\label{sec:3}}
In quantum mechanics, we have the  de~Broglie relations:
\begin{eqnarray}
\lambda = h/p;\\
 \nu = E/h.
\end{eqnarray}
In the previous section we arrive at the conclusion that space-time
must be discrete and get the relation of the smallest length of
space:
\begin{equation}
\Delta x \sim \chi l_{\mathrm{P}},
\end{equation}
which here we consider as the smallest uncertain length in quantum
mechanics. The same discussion also applies to $\Delta t$. The
uncertainty of the de~Broglie wave length is almost the same size as
the elementary length $\Delta \lambda \sim \chi l_{\mathrm{P}}$, and
this means that the uncertain momentum of the particle is
\begin{equation}
 \Delta p = h/\Delta\lambda \sim h/\chi l_{\mathrm{P}}.
\end{equation}
With these two relations, we immediately get the Heisenberg
uncertainty relation
\begin{equation}
 \Delta x \Delta p \sim h.
\end{equation}

For another uncertainty relation about time and energy, we will
divide it into two different situations. Firstly, we consider the
non-relativistic particle. Using the relation implied by
Badiali~\cite{ETS},
\begin{equation}
\Delta p = m \Delta x/ {\Delta t},
\end{equation}
we have
\begin{equation}
\Delta E = (\Delta p)^2/2m = \Delta p \Delta x/{2\Delta t}.
\end{equation}
Thus we get
 \begin{equation}
 \Delta t \Delta E \sim \Delta x \Delta p/2 \sim h/2.
 \end{equation}
Otherwise, in the relativistic situation, there is a little
difference:
\begin{eqnarray}
\Delta E = \Delta p\ c;\ \\
\Delta t \Delta E \sim \Delta x \Delta p \sim h.
\end{eqnarray}
To sum up, unlike the explanation in Ref.~\cite{ETS}, where Badiali
demonstrated the uncertainty principle from the discreteness of
space-time and the set-up $(\Delta x)^2/ {\Delta t} = \hbar/m$
issued from string theory, we get the two uncertainty relations
$\Delta x \Delta p \sim h$ and $\Delta t \Delta E \sim h$ only from
the discrete space-time derived in the above section. That means
that quantum mechanics, especially the Heisenberg uncertainty
principle is related closely with the discreteness of space-time
structure. In the above discussion, we try to offer an understanding
of the uncertainty principle from the viewpoint of discrete
space-time, rather than a concrete proof.

\section{Information and black hole entropy\label{sec:4}}
The connection between information and entropy is well known. The
entropy of a system measures the uncertainty or lack of information
about the internal configuration of the system. Suppose all we know
about a system are the probabilities $p_n$ corresponding to the
$n$th state of the system. Then according to Shannon's entropy
formula~\cite{MTC}, the entropy associated with the system is:
\begin{equation}
S = - \sum_{n} p_n \ln p_n.
\end{equation}
The unit of information is bit, which is the information available
when the answer to the yes-or-no question is known. So, according to
the previous formula, the entropy of the information is maximized
when the probability amplitude is $p_{\mathrm{yes}} =
p_{\mathrm{no}} = {1}/{2}$. Thus, the entropy of one bit information
is $\ln 2$.

It is widely believed that the information of an elementary system
is one bit. Suppose the minimum volume of space can only accommodate
the simplest binary system. That is to say that the maximum
accommodation of information of the minimum volume is one bit. From
a practical viewpoint, it is well-known that there is no pure plane
without thickness, so the event horizon can be considered as a thin
layer with $\Delta R = \chi l_{\mathrm{P}}$ as its thickness. Hence,
the number of elementary systems is:
\begin{equation}
N = \frac{4 \pi R^2 \Delta R}{ (\chi l_{\mathrm{P}})^3} = \frac{4\pi
R^2}{(\chi l_{\mathrm{P}})^2}.
\end{equation}
The entropy of one bit information is $\ln 2$. Multiplied with the
unit of entropy $k_{\mathrm{B}}$, the entropy of the event horizon
is:
\begin{equation}
S = N k_{\mathrm{B}}\ln2 = \frac{k_{\mathrm{B}} A \ln2}{(\chi
l_{\mathrm{P}})^2}. \label{entropyEH}
\end{equation}
The entropy of the black hole to the outside observer is embodied in
the event horizon and its value is $S = {k_{\mathrm{B}}
A}/{4l_{\mathrm{P}}^2}$. Substituting the equation to
Eq.~(\ref{entropyEH}), we obtain the numerical value of $\chi$:
\begin{equation}
\chi = 2 \sqrt{\ln 2},
\end{equation}
which is consistent with the relation we get above: $\chi \geq 0.1$.
Therefore, the accurate length of the minimum space is $2\sqrt{\ln
2} l_{\mathrm{P}} = 2.7\times 10^{-35}$~m, and the elementary area
is $A_0 = 4{\ln2} \,{l_{\mathrm{P}}^2}$, just the same as implied by
't~Hooft in Refs.~\cite{DRQG,THP}. In Ref.~\cite{NFL}, it is
suggested that a length scale $l = l_{\mathrm{P}}/\eta$ might
replace the gravity constant $G$ as a fundamental unit. Here we give
a precise value of the parameter $\eta=1/\chi= {1}/{2\sqrt{\ln 2}}$.
From that one can get a quantum gravity energy scale $E=\eta
E_{\mathrm{P}}$ where $E_{\rm P} \equiv \sqrt{\hbar c^5/G}$ is the
Planck energy. Such an energy scale is consistent with the gamma-ray
burst observation~\cite{fermi09b} concerning the constraint on the
Lorentz-violation scale~\cite{xm09,sxm10}.

From section 2, we know that the highest temperature the black body
can have is $T\leq \Big({135}/{16{\pi}^2 R}\Big)^{\frac{1}{3}} \leq
1.1\times 10^{32}$~K. Beyond this temperature, the discrete
space-time structure might fuse and the physics there is unknown to
us.

On the black hole event horizon, every elementary volume is an
elementarily binary system, and this demonstrates the saturation of
the entropy bound of the black hole. All these elementarily binary
systems are fully quantized. It is natural to ask why other ordinary
systems can not saturate the entropy bound? That may be because the
elementary units on their boundaries are not fully quantized. This
indicates that not all of them are binary systems, and it may
explain why the entropy of an ordinary system is less than the
entropy bound.

\section{Comments and Summary\label{sec:5}}

In fact, the idea of space-time discreteness is not new, and there
have been many arguments and debates from philosophical viewpoint in
human history. Many discussions are well-known to us, such as Zeno's
paradoxes, but most of them belong to the domain of metephysics. An
origin of the discreteness of space-time from a physical viewpoint
might be traced back to Planck, when he constructed a basic unit
system in 1899~\cite{p99} before his idea of energy quanta, {\it
i.e.}, the birth of quantum theory in 1990. In Planck's unit system,
there are five basic physical quantities, {\it i.e.}, the light
speed $c$, the gravitational constant $G$, the Boltzmann constant
$k_B$, and $1/4\pi \epsilon_0$ (where $\epsilon_0$ is the
permittivity of free space or the electric constant), together with
the new quantity $\hbar$ which is well-known as the Planck constant
in quantum theory. There are a number of basic quantities in this
unit system, such as the Planck length $l_{\rm P} \equiv
\sqrt{G\hbar/c^3} \simeq 1.6 \times 10^{-35}$~m, the Planck time
$t_{\rm P} \equiv \sqrt{G\hbar/c^5} \simeq 5.4 \times 10^{-44}$~s,
the Planck energy $E_{\rm P} \equiv \sqrt{\hbar c^5/G} \simeq 2.0
\times 10^{9}$~J, and the Planck temperature $T_{\rm P} \equiv
\sqrt{\hbar c^5/Gk_B^2} \simeq 1.4 \times 10^{32}$~K.

As energy is quantized in quantum theory, the idea for an extension
to the quantization of space and time has been discussed by many
physicists afterward. There are have been many theories in which the
concept of discreteness of space-time has been implemented, such as
the non-commutative space-time~\cite{Snyder}. Some developments of
quantum gravity (QG) adopted a smallest length scale for the
structure of space-time, or equivalently, an upper energy bound for
particles in the quantum geometrical
background~\cite{AmelinoCamelia:2000ge,AmelinoCamelia:2000mn,KowalskiGlikman:2001gp,KowalskiGlikman:2001ct,Magueijo:2001cr,Magueijo:2002am,AmelinoCamelia:2002vy,AmelinoCamelia:2010pd}.
The most natural candidate for the minimal length appears to be the
Planck length $l_{\rm P}$, or accordingly, the Planck energy $E_{\rm
P}$, for the maximal energy of particles. However, there have been
suggestions~\cite{NFL} on the presence of a new fundamental length
scale $l$, which, from dimensional analysis, is proportional to
$l_{\rm P}$, \textit{i.e.}, $l = l_{\rm P}/\eta$, where $\eta$
corresponds to $1/\chi={1}/{2\sqrt{\ln 2}}$ as pointed out in
Section~\ref{sec:4} of this paper.

Though a minimal length of space has been formulated in many
theories, it is usually taken for granted as a hypothesis without
physical explanation from a more concrete foundation. In this
article, by applying the generalized second law and two universal
entropy bounds suggested by Susskind and Bekenstein to the
black-body radiation, we arrive at a minimum length scale of space.
We also get the minimum time interval. Therefore, we reveal from
physical arguments that space-time is discrete rather than
continuous. Space and time can not be infinitesimal, but have
elementary units.

We can explain the Heisenberg uncertainty relation using the
discreteness of space-time. We also get the precise value of the
elementary length. The highest temperature under which the
space-time still holds is also obtained. The minimum length scale we
got is a little different from the Planck length. It is a little
larger than the Planck length many people thought before. Here we
give the acquired minimum length scale more physically and
concretely. The existence of a smallest length scale can manifest
itself through the physical consequence of Lorentz invariance
violation, which is a frontier under extensive investigations both
theoretically and experimentally~\cite{ShaoMaReview}. Thus the idea
of discreteness of space-time can be tested through high precision
measurements of Lorentz invariance violation effects.

From information theory, we can interpret the universal entropy
bound and explain the inconsistency of entropies between black holes
and ordinary systems. It is expected that fruitful results can be
got from the new understanding of space-time discreteness.

\noindent{\bf Acknowledgments} We thank Zhi Xiao and Lijing Shao for
helpful discussions. This work is partially supported by National
Natural Science Foundation of China (Grants No.~11021092,
No.~10975003, No.~11035003).



\vspace{2pc}
\begin{thebibliography}{10}


\bibitem{Hawking1}
S.W. Hawking, Phys. Rev. Lett. 26, 1344 (1971).
\bibitem{Hawking2}
S.W. Hawking, Comm. Math. Phys. 25, 152 (1972).
\bibitem{BHSL} J.D. Bekenstein, \Journal{Nuovo Cim.
Lett.}{4}{1972}{737}.
\bibitem{BHE} J.D. Bekenstein, \Journal{Phys. Rev.
D}{7}{1973}{2333}.
\bibitem{GSL} J.D. Bekenstein, \Journal{Phys. Rev.
D}{9}{1974}{3292}.
\bibitem{BHEH} S.W. Hawking, \Journal{Nature}{30}{1974}{248}.
\bibitem{PCBH} S.W. Hawking, \Journal{Commun. Math.
Phys.}{43}{1975}{199}.
\bibitem{UUB} J.D. Bekenstein, \Journal{Phys. Rev.
D}{23}{1981}{287}.
\bibitem{TWAH} L. Susskind, \Journal{J. Math. Phys.}{36}{1995}{6377} [arXiv:
hep-th/9409089].
\bibitem{THPB} R. Bousso, \Journal{Rev. Mod. Phys.}{74}{2002}{825} [arXiv:
hep-th/0203101].
\bibitem{ETS} J.P. Badiali, \Journal{J. Phys. A}{38}{2005}{2835} [arXiv:
quant-ph/0409138].
\bibitem{MTC} C.E. Shannon and W. Weaver, The Mathematical Theory of
Communication. University of Illinois Press, Urbana (Ill.), 1949.
\bibitem{DRQG} G. 't Hooft, arXiv: gr-qc/9310026.
\bibitem{THP} G. 't Hooft, arXiv: hep-th/0003004.
\bibitem{NFL} For a brief discussion and review on a minimal lenghth scale, see, e.g., L.~Shao, B.-Q.~Ma, arXiv: hep-th/1006.3031.
\bibitem{fermi09b}
A.A. Abdo {\it et al.}, \Journal{Nature}{462}{2009}{331}
[arXiv:0908.1832].

\bibitem{xm09}
Z. Xiao, B.-Q. Ma, \Journal{\PRD}{80}{2009}{116005}
[arXiv:0909.4927].

\bibitem{sxm10}
L. Shao, Z. Xiao, B.-Q. Ma, \Journal{\ApP}{33}{2010}{312}.
arXiv:0911.2276.

\bibitem{p99}
M. Planck, \Journal{\SKPAWB}{5}{1899}{440}.

\bibitem{Snyder}
H.S. Snyder, Phys. Rev. 71 , 38 (1947); 72, 68 (1947).

\bibitem{AmelinoCamelia:2000ge}
  G.~Amelino-Camelia,
  Phys.\ Lett.\  B {\bf 510}, 255 (2001)
  [arXiv:hep-th/0012238].

\bibitem{AmelinoCamelia:2000mn}
  G.~Amelino-Camelia,
  Int.\ J.\ Mod.\ Phys.\  D {\bf 11}, 35 (2002)
  [arXiv:gr-qc/0012051].

\bibitem{KowalskiGlikman:2001gp}
  J.~Kowalski-Glikman,
  Phys.\ Lett.\  A {\bf 286}, 391 (2001)
  [arXiv:hep-th/0102098].


\bibitem{KowalskiGlikman:2001ct}
  J.~Kowalski-Glikman,
  Phys.\ Lett.\  A {\bf 299}, 454 (2002)
  [arXiv:hep-th/0111110].

\bibitem{Magueijo:2001cr}
  J.~Magueijo and L.~Smolin,
  Phys.\ Rev.\ Lett.\  {\bf 88}, 190403 (2002)
  [arXiv:hep-th/0112090].

\bibitem{Magueijo:2002am}
  J.~Magueijo and L.~Smolin,
  Phys.\ Rev.\  D {\bf 67}, 044017 (2003)
  [arXiv:gr-qc/0207085].

\bibitem{AmelinoCamelia:2002vy}
  G.~Amelino-Camelia,
  Int.\ J.\ Mod.\ Phys.\  D {\bf 11}, 1643 (2002)
  [arXiv:gr-qc/0210063].

\bibitem{AmelinoCamelia:2010pd}
  G.~Amelino-Camelia,
  Symmetry {\bf 2}, 230 (2010)
  [arXiv:1003.3942 [gr-qc]].


\bibitem{ShaoMaReview}
For a brief review on Lorentz violation effects through very high
energy photons of astrophysical sources, see. e.g.,
  L.~Shao and B.-Q.~Ma,
  Mod.\ Phys.\ Lett.\  A {\bf 25}, 3251 (2010)
  [arXiv:1007.2269 [hep-ph]].



\end{thebibliography}
\end{document}